\documentclass[aps,prb,twocolumn]{revtex4-1}
\usepackage{amssymb}
\usepackage{graphicx}

\begin{document}
\author{Kavita Mehlawat$^1$, A. Thamizhavel$^2$ and Yogesh Singh$^1$ }
\affiliation{$^1$Department of Physical Sciences, Indian Institute of Science Education and Research (IISER) Mohali, Knowledge City, Sector 81, Mohali 140306, India.\\
$^2$ Department of Condensed Matter Physics and Material Sciences, Tata Institute of Fundamental Research, Mumbai 400005, India.}

\date{\today}

\title{Heat capacity evidence for proximity to the Kitaev QSL in A$_2$IrO$_3$ ($A =$ Na, Li)}

\begin{abstract}
The honeycomb lattice iridates $A_2$IrO$_3$ ($A =$ Na, Li) are candidates for realization of the Kitaev-Heisenberg model although their proximity to Kitaev's quantum Spin-Liquid (QSL) is still debated.  We report on heat capacity $C$ and entropy $S_{mag}$ for $A_2$IrO$_3$ ($A =$ Na, Li) in the temperature range $0.075~{\rm K}\leq T \leq 155$~K\@.  We find a well separated two-peak structure for the magnetic heat capacity $C_{mag}$ for both materials and $S_{mag}$ for Na$_2$IrO$_3$ shows a shoulder between the peaks with a value close to ${1\over 2}$Rln$2$.  These features signal the fractionalization of spins into Majornana Fermions close to Kitaev's QSL [Phys. Rev. B {\bf 92}, 115122 (2015); Phys. Rev. B {\bf 93}, 174425 (2016).]. These results provide the first thermodynamic evidence that $A_2$IrO$_3$ are situated close to the Kitaev QSL.  Additionally we measure the high temperature $T\leq 1000$~K magnetic susceptibility $\chi$ and estimate the Weiss temperature $\theta$ in the true paramagnetic state.  We find $\theta \approx -127$~K and $-105$~K, for Na$_2$IrO$_3$ and Li$_2$IrO$_3$, respectively.
 
\end{abstract}

\maketitle

The family of layered honeycomb lattice iridates $A_{2}$IrO$_{3} (A =$~Na, Li) has garnered a lot of recent attention as possible realizations of the Kitaev-Heisenberg model \cite{Jackeli2009, Chaloupka2010, Singh2010, Liu2011, Kimchi2011, Choi2012, Ye2012, Singh2012, Gretarsson2012, Comin2012, Chaloupka2013, Gretarsson2013, Foyevtsova2013, Katukuri2014, Yamaji2014, Sizyuk2014}.  Although these materials were found to be magnetically ordered at low temperatures \cite{Singh2010, Choi2012, Ye2012, Singh2012} there is growing evidence of dominant Kitaev interactions from $ab ~initio$ estimations of the exchange parameters \cite{Foyevtsova2013, Katukuri2014,Yamaji2014}, from inelastic Raman scattering measurements \cite{Gupta2016}, and from direct evidence of dominant bond directional exchange interactions in Na$_{2}$IrO$ _{3}$\cite{Chun2015}.  It is still debated how close or far the real materials are from the Quantum Spin Liquid (QSL) expected in the strong Kitaev limit.  Recently predictions for signatures of Kitaev's QSL in Raman scattering experiments have been made \cite{Knolle2014} and have been observed in experiments on Na$_2$IrO$_3$ and (Na$_{1-x}$Li$_x$)$_2$IrO$_3$  \cite{Gupta2016}. Optical spectroscopy measurements on Na$_2$IrO$_3$ also claimed to observe signatures consistent with proximity to a QSL state \cite{Alpichshev2015}.  However, direct thermodynamic evidence is still lacking due to the unavailability of clear criteria applicable to the real materials.  

Recently, thermodynamic properties of the Kitaev model on a honeycomb lattice have been calculated and it was shown that heat capacity would show a two-peak structure coming from the fractionalization of the quantum spins into two kinds (dispersive and dispersionless) of Majorana Fermions \cite{Nasu2015}.  The two peaks in the heat capacity would then come from the thermal excitation of these two kinds of Majorana Fermions.  The Rln$2$ entropy of the spins would be shared equally between the two Majorana's and so the temperature dependence of the magnetic entropy $S_{mag}$ would show a half-plateau between the two heat capacity peaks and the value of $S_{mag}$ at the plateau would be pinned to ${1\over 2}$Rln$2$ ~\cite{Nasu2015}.  These predictions for the Kitaev model are however, still not applicable to the real materials which are magnetically ordered \cite{Singh2010, Ye2012, Singh2012} and have terms other than the dominant Kitaev term present in their Hamiltonian \cite{Foyevtsova2013, Katukuri2014,Yamaji2014}.  More recently, thermodynamic properties have been determined for a generalized Kitaev-Heisenberg model and for the $ab~initio$ Hamiltonian for Na$_2$IrO$_3$ arrived at in Ref.~\onlinecite{Yamaji2014}.  It was predicted that even for magnetically ordered material proximate to Kitaev's QSL the two-peak structure in heat capacity would survive while the plateau in $S_{mag}$ at ${1\over 2}$Rln$2$ becomes a shoulder at the same numerical value\cite{Yamaji2016}.  These features vanish when the material is deep into the magnetically ordered state away from the QSL state \cite{Yamaji2016}.  Thus, a well separated two-peak structure in the magnetic heat capacity and a shoulder in $S_{mag}$ between the two peaks with a value close to Rln$2$ is the predicted hallmark to place $A_2$IrO$_3$ materials proximate to Kitaev's QSL \cite{Yamaji2016}.   


In this work we report high temperature ($T \leq 1000$~K) magnetic susceptibility $\chi$ versus temperature and heat capacity measurements in the $T$ range $0.075~{\rm K}\leq T \leq 155$~K on polycrystalline samples of the honeycomb lattice iridates $A_2$IrO$_3$.  Our measurements provide three new results. (i) The high temperature $\chi(T)$ data gave the Weiss temperatures $\theta = -127(4)$~K and $-105(2)$~K for Na$_2$IrO$_3$ and Li$_2$IrO$_3$, respectively.  While $\theta$ for Na$_2$IrO$_3$ is similar to values found previously using lower temperature ($T < 300$~K) $\chi(T)$ data \cite{Singh2010, Singh2012}, $\theta$ for Li$_2$IrO$_3$ is larger by a factor of $3$ compared to the values found using lower temperature $\chi(T)$ data \cite{Singh2012}.  This indicates that, in contrast to what was believed previously, magnetic energy scales in both materials should be similar.  (ii) The magnetic contribution to heat capacity shows a two-peak structure and the $T$ dependence of the entropy $S_{mag}$ shows a shoulder between the two peaks with a value close to ${1\over 2}$Rln$2$ for Na$_2$IrO$_3$.  This is in excellent agreement with recent predictions \cite{Nasu2015,Yamaji2016} and provides the first thermodynamic evidence that Na$_2$IrO$_3$ lies proximate to Kitaev's QSL.  The results for Li$_2$IrO$_3$ are qualitatively similar although quantitative agreement is not as strong. (iii) Lastly, the low temperature $C$ for Na$_2$IrO$_3$ and Li$_2$IrO$_3$ show very different $T$ dependence.  While $C(T)$ for Na$_2$IrO$_3$ shows a conventional $T^3$ behaviour, $C(T)$ for Li$_2$IrO$_3$ shows a clear $T^2$ dependence suggesting novel $2$-dimensional magnetic excitations for this material.  

\noindent
\emph{Experimental:}
Polycrystalline samples of $A_2$IrO$_3$ were synthesized using a solid state reaction method starting with high purity chemicals and heating the pelletized mixtures between $900$ and $1000~^o$C in $50~^0$C steps.  The step-wise heating instead of going directly to $1000^o$~C was found to be essential for the formation of high quality samples.  Powder X-ray diffraction on crushed pieces of the samples confirmed the formation of single phase samples with the correct lattice parameters \cite{Singh2012}.  

The magnetic susceptibility $\chi$ versus temperature data were measured in the temperature range $T = 2$~K to $400$~K using the VSM option of a physical property measurement system from Quantum Design (QD-PPMS).  The $\chi(T)$ data between $T = 300$~K and $1000$~K were measured using the VSM oven option of the QD-PPMS.  The heat capacity $C$ data were measured in the temperature range $2$~K to $155$~K using a QD-PPMS.  The $C$ data from $75$~mK to $3$~K was measured using the dilution refrigerator (DR) option of a QD-PPMS.

\begin{figure}[t]   
\includegraphics[width= 3 in]{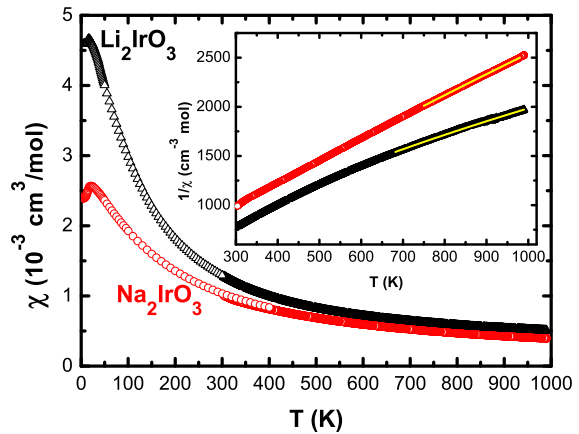}    
\caption{(Color online) Magnetic susceptibility $\chi$ versus temperature $T$ for $A_2$IrO$_3$ ($A =$ Na, Li) measured between $T = 2$~K and $1000$~K in a magnetic field of $H = 1$~T\@.  Inset shows the $1/\chi(T)$ data above $T = 300$~K\@.  The solid curves through the data are fits of the high temperature data to the Curie-Weiss expression (see text for details).  
\label{Fig-chi}}
\end{figure}

\noindent
\emph{Magnetic Susceptibility:}
Figure~\ref{Fig-chi} shows the $\chi$ versus $T$ data for $A_2$IrO$_3$ between $2$ and $1000$~K\@.  The two separate measurements, between $2$ and $400$~K and between $300$ and $1000$~K, for each sample match quite well.  Sharp anomalies were observed at $T_N \approx 15$~K for both samples in agreement with previous reports \cite{Singh2012}.  The new data are the ones above $T = 400$~K\@.  These data are plotted as $1/\chi(T)$ in the inset of Fig.~\ref{Fig-chi}.  Data for Na$_2$IrO$_3$ are approximately linear in this temperature range.  The data above $T \approx 750$~K were fit by the expression $\chi(T) = \chi_0+{C\over T-\theta}$, with $\chi_0$ , $C$, and $\theta$ as fit parameters.  Here $\chi_0$ is a $T$ independent term, $C$ is the Curie constant, and $\theta$ is the Weiss temperature.  The fit gave the values $\chi_0 = 2.66(3) \times 10^{-5}$~cm$^3$/mol, $C = 0.395(1)$~cm$^3$~K/mol, and $\theta = -127(4)$~K\@.  These values are close to those found by fitting the $\chi(T)$ data for $T \leq 300$~K  \cite{Singh2010,Singh2012}.  In particular, the value of $\theta$, which gives the overall scale of the magnetic interactions, comes out to be very close to the value $-125(6)$~K found previously for polycrystalline Na$_2$IrO$_3$ \cite{Singh2012}.   
 
For Li$_2$IrO$_3$ the $1/\chi(T)$ data shown in the inset of Fig.~\ref{Fig-chi} shows a prominent downward curvature.  This suggests a large and positive $\chi_0$.  The large curvature also means that to obtain a reliable value of $\theta$ one needs to be well in the paramagnetic state.  The fit to the data above $T = 700$~K gave the values $\chi_0 = 1.45(3) \times 10^{-4}$~cm$^3$/mol, $C = 0.403(2)$~cm$^3$~K/mol, and $\theta = -105(2)$~K\@.  The values of $\chi_0$ is slightly larger than obtained previously.  This suggests a large Van Vleck paramagnetic contribution for Li$_2$IrO$_3$.  The most conspicuous difference between the low temperature and high temperature fit parameters is the value of $\theta = -105(2)$~K which is about a factor of $3$ larger in magnitude compared to the value $-33$~K obtained previously \cite{Singh2012}.  This indicates that, in contrast to what was believed previously based on old $\theta$ values, magnetic energy scales in both materials might be similar.

\begin{figure}[t]   
\includegraphics[width= 3 in]{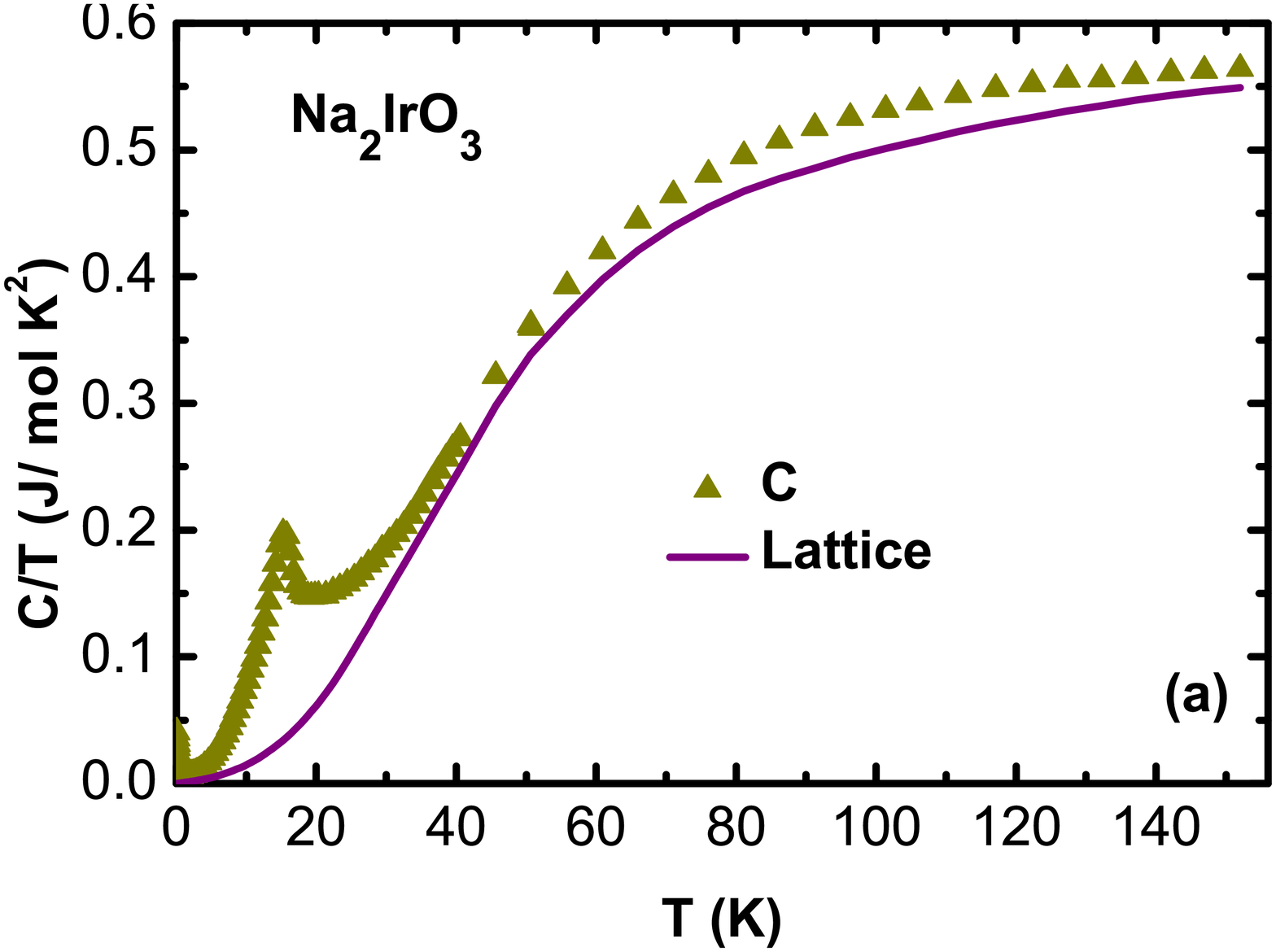}    
\includegraphics[width= 3 in]{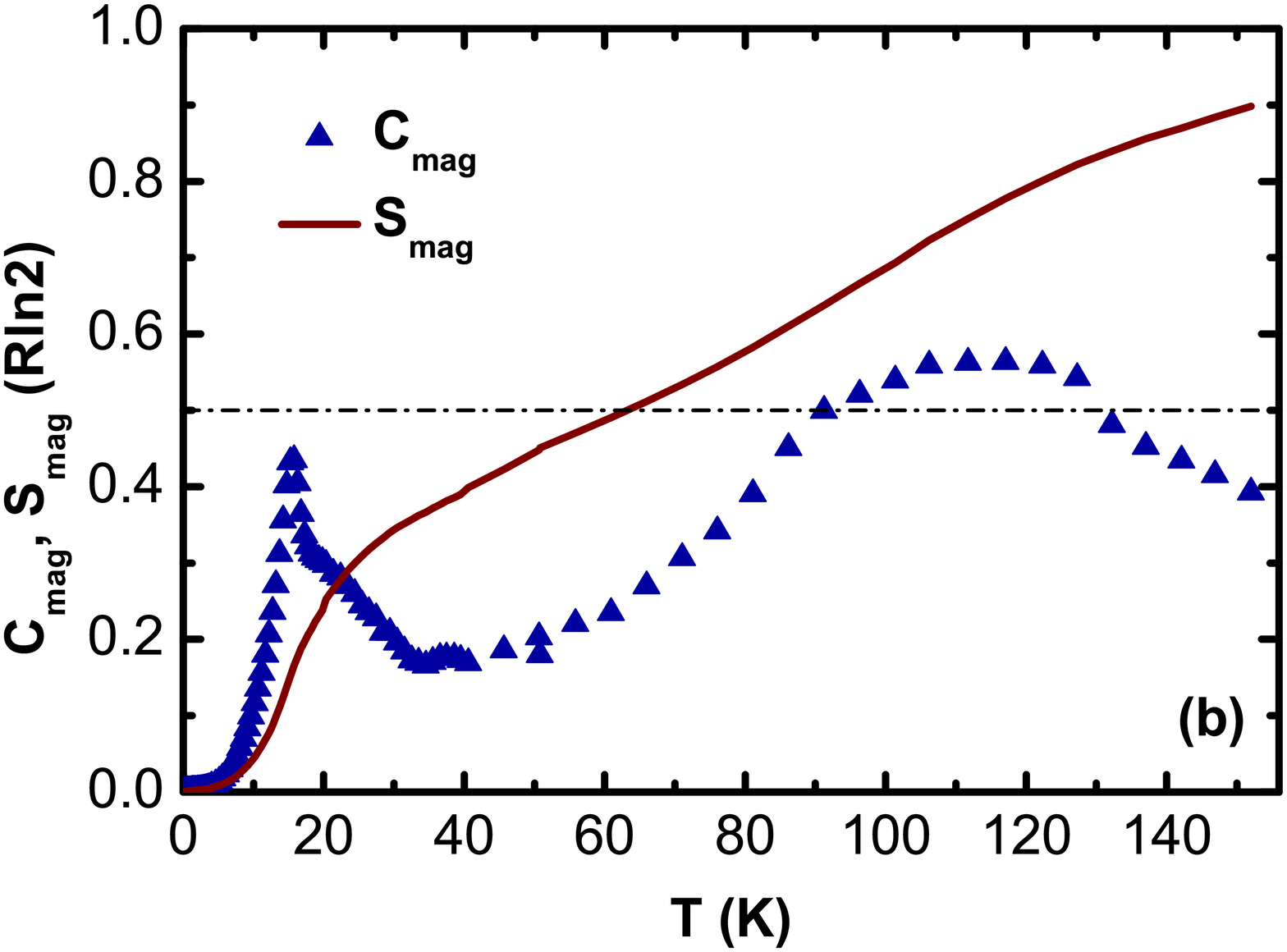}    
\caption{(Color online) (a) Heat capacity divided by temperature $C/T$ versus $T$ for Na$_2$IrO$_3$.  The lattice contribution to the heat capacity is shown as the solid curve.  (b) Magnetic contribution to the heat capacity $C_{mag}$ and the magnetic entropy $S_{mag}$ in units of Rln$2$ versus $T$ for Na$_2$IrO$_3$.  The horizontal dash-dot line is the value ${1\over 2}$Rln$2$.    
\label{Fig-NaIrO}}
\end{figure}
\noindent
\emph{Heat Capacity:}  Figure~\ref{Fig-NaIrO}~(a) show the heat capacity divided by temperature $C/T$ versus $T$ data for Na$_2$IrO$_3$ between $T = 75$~mK and $155$~K\@.  A sharp $\lambda$~type anomaly near $15$~K confirms the antiferromagnetic transition for Na$_2$IrO$_3$ \cite{Singh2010, Singh2012}.  This anomaly is much sharper than observed for previous polycrystalline samples \cite{Singh2012} indicating the high quality of the sample used in the current study.  The data at lower temperatures show an upturn below about $1$~K which we will return to later.  We also show in Fig~\ref{Fig-NaIrO}~(a) the approximate lattice contribution obtained by measuring the heat capacity of the iso-structural non-magnetic analog Na$_2$SnO$_3$ and then rescaling the data to account for the molecular mass difference between Na$_2$IrO$_3$ and Na$_2$SnO$_3$.  By subtracting this lattice contribution from the total $C(T)$ one can obtain the magnetic contribution $C_{mag}$  shown (in the units Rln$2$) in Fig.~\ref{Fig-NaIrO}~(b).  In addition to the low temperature anomaly we find another broad peak centered around $\approx 110$~K\@.  Such a two-peak structure has been predicted recently for a generalized Kitaev-Heisenberg Hamiltonian for parameters placing the material in the magnetic state proximate to Kitaev's QSL \cite{Yamaji2016}.  A two-peak structure is however, not uncommon in frustrated and/or low-dimensional magnetic materials where the high temperature anomaly occurs when short ranged magnetic correlations start to develop while the low temperature peak occurs on the development of long ranged correlations \cite{Hardy2003}.  However, the definitive signature for closeness to Kitaev's QSL has been predicted to be the $T$ dependence of the entropy $S_{mag}$ which must show a half-plateau pinned at or close to the value ${1\over 2}$Rln$2$ between the two heat capacity peaks~ \cite{Nasu2015, Yamaji2016}.      

\begin{figure}[t]   
\includegraphics[width= 3 in]{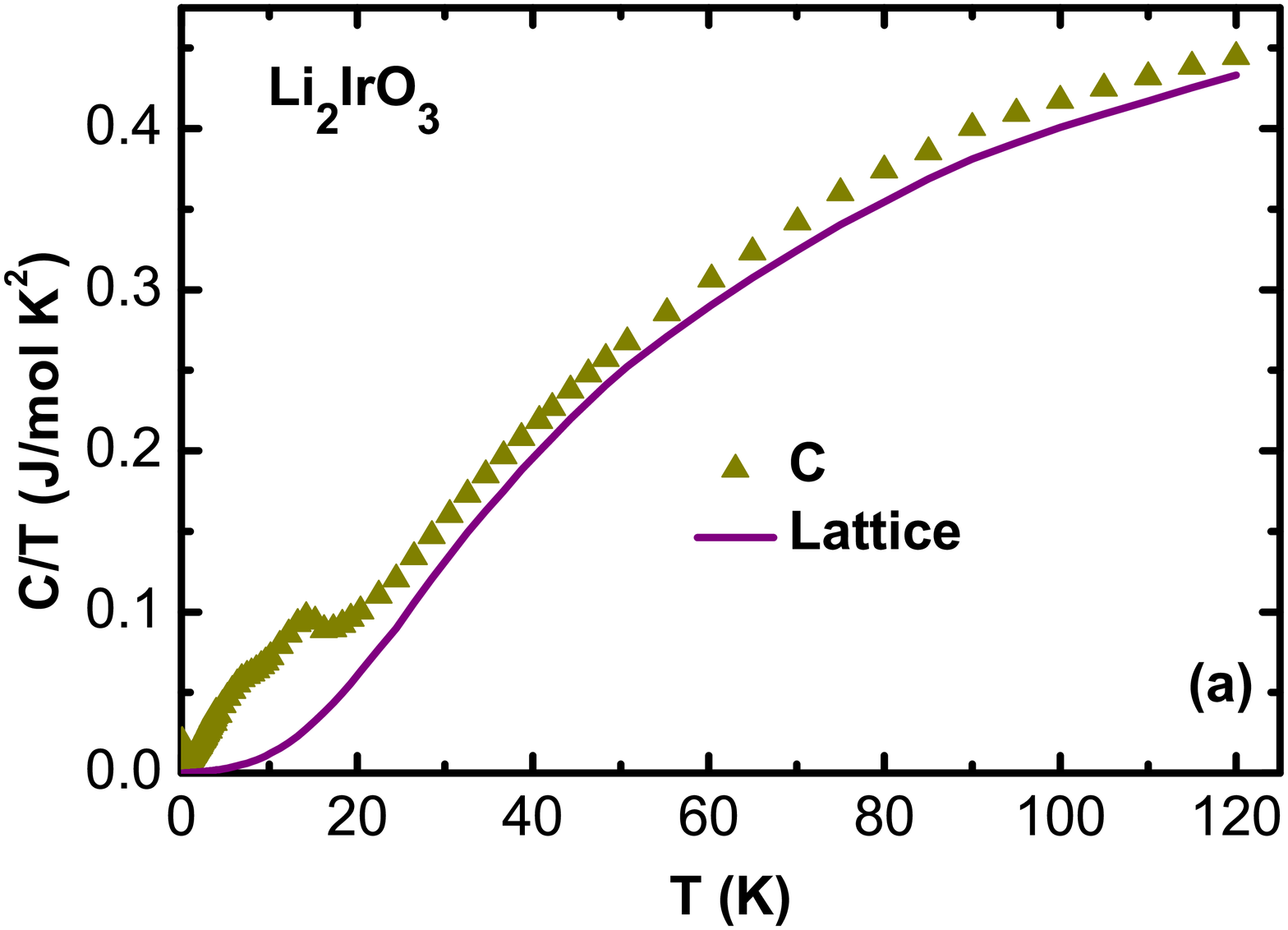}    
\includegraphics[width= 3 in]{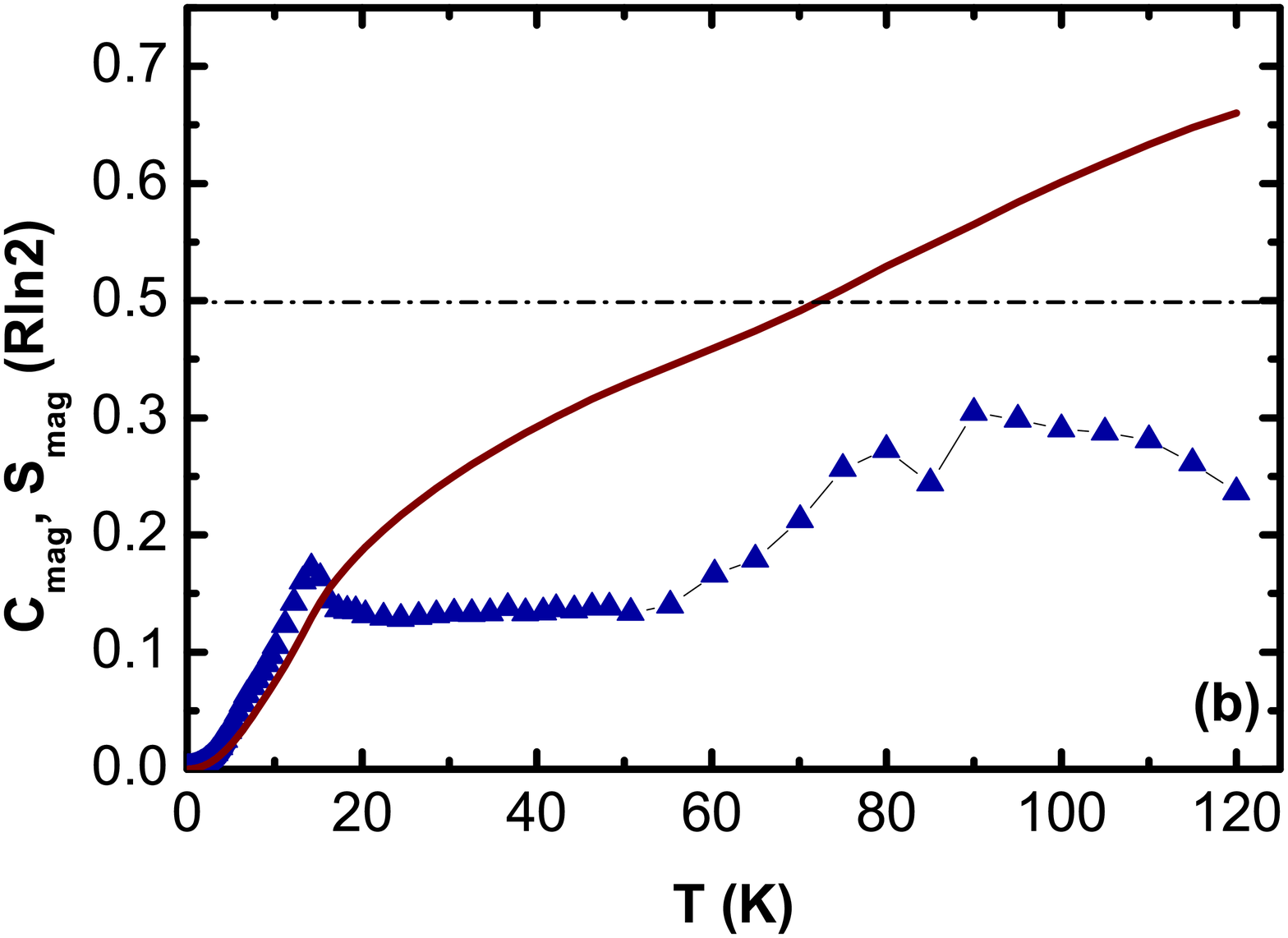}    
\caption{(Color online) (a) Heat capacity divided by temperature $C/T$ versus $T$ for Li$_2$IrO$_3$.  The lattice contribution to the heat capacity is shown as the solid curve.  (b) Magnetic contribution to the heat capacity $C_{mag}$ and the magnetic entropy $S_{mag}$ in units of Rln$2$ versus $T$ for Li$_2$IrO$_3$.  The horizontal dash-dot line is the value ${1\over 2}$Rln$2$.   
\label{Fig-LiIrO}}
\end{figure}

We present the $T$ dependence of $S_{mag}$ in units of Rln$2$ in Fig.~\ref{Fig-NaIrO}~(b).  We note that there is a distinct shoulder in $S_{mag}(T)$ between the two heat capacity peaks and that the value of $S_{mag}$ between the peaks is close to the predicted value ${1\over 2}$Rln$2$ shown as the horizontal dashed line in Fig.~\ref{Fig-NaIrO}~(b).  $S_{mag}$ reaches $90\%$Rln$2$ at the highest $T$ of our measurements.  The high temperature peak in $C_{mag}$ is quite broad and one can see a tail extending to even higher temperatures.  It is evident that the full Rln$2$ entropy will be recovered at a slightly higher temperature.  The two-peak structure in heat capacity, the $T$ dependence of $S_{mag}$ with a half-plateau between the peaks and its numerical value $={1\over 2}$Rln$2$ are in excellent agreement with theoretical predictions and provide direct evidence which lead to the inference that Na$_2$IrO$_3$ is situated close to Kitaev's QSL.   

We now turn to heat capacity data on Li$_2$IrO$_3$.  The $C/T$ versus $T$ data for Li$_2$IrO$_3$ are shown in Fig.~\ref{Fig-LiIrO}~(a) between $T = 90$~mK and $120$~K\@.  The lattice contribution, estimated by measuring the heat capacity of the isostructural non-magnetic material Li$_2$SnO$_3$ and rescaling the data to account for the difference in molecular masses of Li$_2$IrO$_3$ and Li$_2$SnO$_3$, is also shown in Fig.~\ref{Fig-LiIrO}~(a).  The $15$~K anomaly signalling the onset of long-ranged zig-zag magnetic order is clearly visible as is a weak shoulder around $7$~K\@.  This shoulder below the main magnetic anomaly has been observed for all previous polycrystalline samples as well \cite{Singh2012, Manni-thesis}.  This second anomaly is most likely associated with disorder as its relative magnitude compared to the $15$~K anomaly can be suppressed by improving the quality of the samples \cite{Manni-thesis}.  It must be noted however that the second anomaly cannot be completely suppressed even for the best samples (including single crystals \cite{Freund2016}) and our current sample is at least as good as the best polycrystalline samples produced thus far \cite{Singh2012,Manni-thesis}.  As for Na$_2$IrO$_3$, the low-$T$ data for Li$_2$IrO$_3$ show an abrupt upturn below about $1$~K\@.  We will discuss the low temperature data for $A_2$IrO$_3$ separately later.      

The $C_{mag}$ data for Li$_2$IrO$_3$  obtained by subtracting the lattice part from the total $C(T)$ is shown in Fig.~\ref{Fig-LiIrO}~(b).  Although there is more scatter in the obtained data compared to Na$_2$IrO$_3$, the two-peak structure in $C_{mag}(T)$ is clearly visible for Li$_2$IrO$_3$ too.  The two peaks occur at $15$~K and $\sim 90$~K, respectively.  The magnetic entropy $S_{mag}$ shown in Fig.~\ref{Fig-LiIrO}~(b) also shows a shoulder between the two peaks although the quantitative match with predictions are not as strong as for Na$_2$IrO$_3$.  Specifically, the $S_{mag}$ value between the two heat capacity peaks reaches only about $35\%$~Rln$2$ and the value ${1\over 2}$Rln$2$ is reached only close to the start of the high temperature peak.  The value of $S_{mag}$ at $120$~K is only about $65\%$Rln$2$ and it seems unlikely that the rest will be recovered under the tail of the high temperature peak beyond $120$~K\@.  The possibility that Li$_2$SnO$_3$ isn't a good approximation for the lattice heat capacity for Li$_2$IrO$_3$ presents itself.  Nevertheless, the $C_{mag}(T)$ data with the two-peak structure and the $S_{mag}(T)$ with a plateau between the two peaks are qualitatively consistent with predictions for materials close to Kitaev's QSL \cite{Yamaji2016}.   

\begin{figure}[t]   
\includegraphics[width= 3 in]{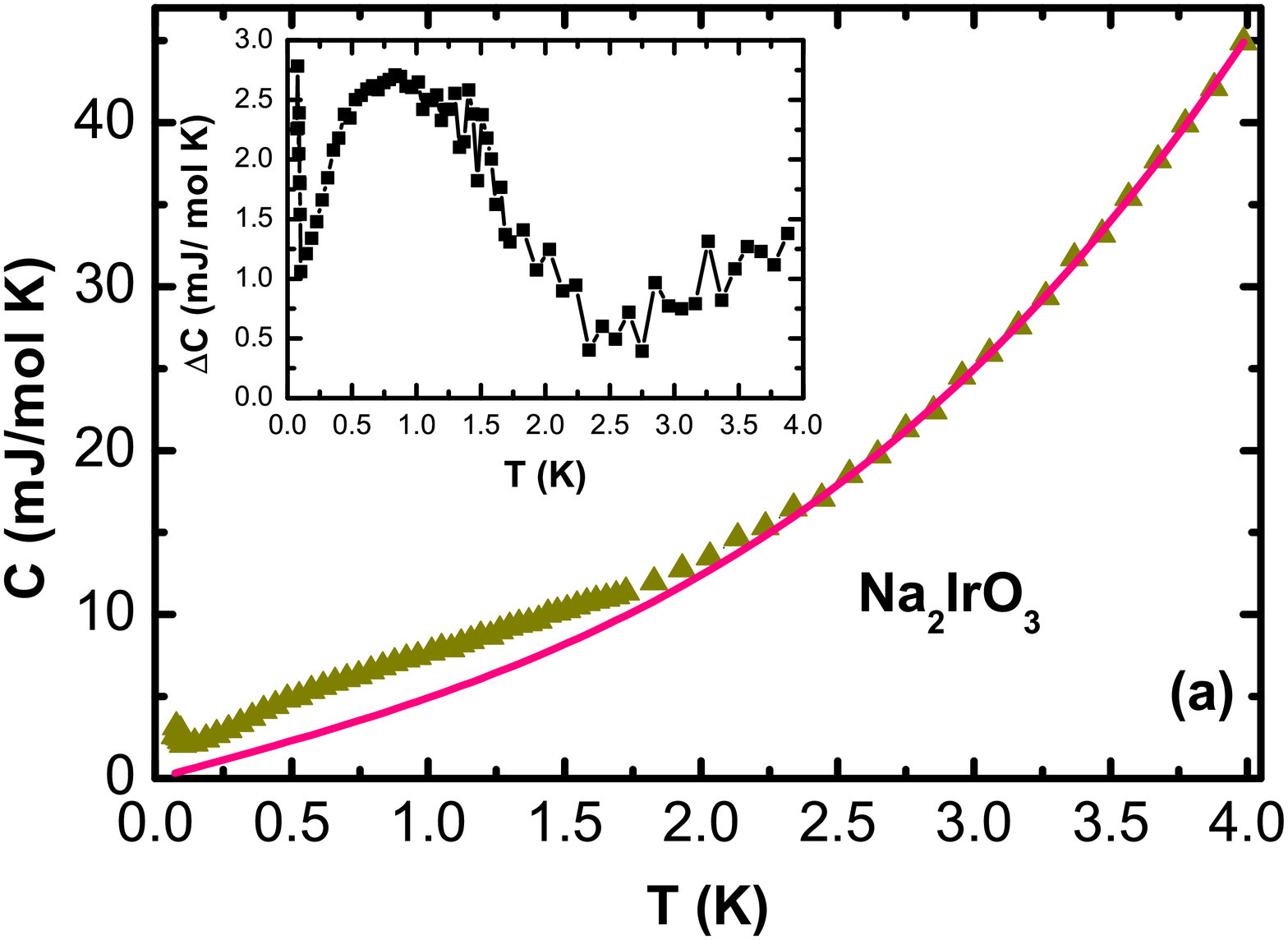}    
\includegraphics[width= 3 in]{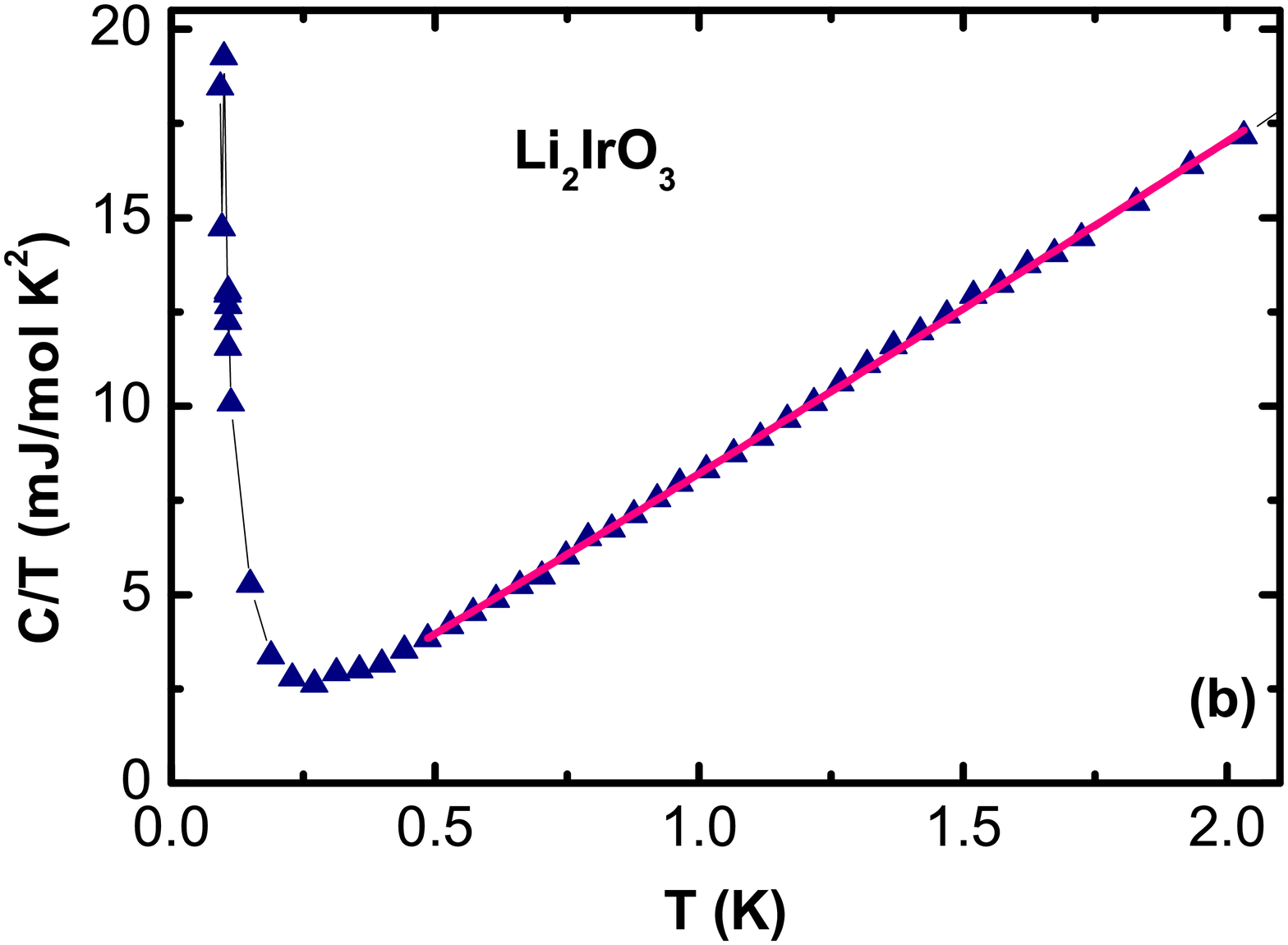}    
\caption{(Color online) (a) Heat capacity $C$ versus temperature $T$ for Na$_2$IrO$_3$ below $T = 4$~K\@.  The solid curve through the data shows the $T^3$ dependence.  The inset shows the weak anomaly around $0.8$~K and the upturn at lower $T$.  (b) $C/T$ vs $T$ for Li$_2$IrO$_3$ below $2$~K\@.  The solid curve through the data shows the $C \sim T^2$ dependence.    
\label{Fig-LT}}
\end{figure}

We finally discuss the low temperature behaviours of $C(T)$ for $A_2$IrO$_3$ and show that they follow qualitatively different $T$ dependences.  Figure~\ref{Fig-LT}~(a) shows the $C$ versus $T$ data for Na$_2$IrO$_3$ measured down to $T = 75$~mK in the DR.  The data between $2$~K and $4$~K follow a $T^3$ behaviour (shown as the solid curve through the data in Fig.~\ref{Fig-LT}~(a))  expected for a $3$-dimensional insulator in the antiferromagnetic state.  In this case the $T^3$ contribution to the heat capacity will be a combination of phonons and antiferromagnetic spin-waves.  From Fig.~\ref{Fig-LT}~(a) it is evident that below $\approx 2$~K there is an excess $C(T)$ above the $T^3$ contribution.  If the $T^3$ contribution is subtracted from $C(T)$ one gets a difference heat capacity $\Delta C(T)$ shown in the inset of Fig.~\ref{Fig-NaIrO}~(a).  A clear anomaly peaked at $\approx 0.8$~K can be observed in the $\Delta C(T)$ data.  The entropy under this peak however is quite small ($\leq 1\%$Rln$2$) suggesting that it could be extrinsic in origin.  The upturn at the lowest temperatures could be the start of a nuclear Schottky anomaly.  

Figure~\ref{Fig-LT}~(b) shows the $C/T$ versus $T$ data for Li$_2$IrO$_3$ measured down to $T = 90$~mK\@.  Below $2$~K the $C/T$ data follow a linear in $T$ behaviour down to about $0.3$~K where an abrupt upturn is observed.  This upturn could again be the high temperature tail of a nuclear Schottky anomaly.  The $C \sim T^2$ behaviour for Li$_2$IrO$_3$ is unusual and suggests different magnetic excitations compared to Na$_2$IrO$_3$ which shows a conventional $C \sim T^3$ behaviour. \\

\noindent
\emph{Summary and Discussion:} $A_2$IrO$_3$ materials have a high magnetic energy scale as suggested by recent Raman \cite{Gupta2016} and RIXS \cite{Chun2015} measurements as well as in $ab~initio$ estimations of the exchange parameters \cite{Foyevtsova2013, Katukuri2014,Yamaji2014}.  This prompted us to measure the high temperature $T \leq 1000$~K magnetic susceptibility and use these to estimate the Weiss temperature $\theta$ in the true paramagnetic state.  This gave the first new result that $\theta$ for Li$_2$IrO$_3$ which was previously estimated to be $\sim -30$~K is actually $\sim -100$~K and much closer to the value for Na$_2$IrO$_3$ indicating that magnetic interaction scales for the two materials are very similar.  
Secondly, motivated by recent theoretical criteria for placing materials close to Kitaev's QSL state \cite{Nasu2015, Yamaji2016}, we have presented the heat capacity $C$ and magnetic entropy $S_{mag}$ from $T =75$~mK to $T > \theta$.  We find a two-peak structure in $C_{mag}$ and $S_{mag}$ shows a clear shoulder with a value ${1\over 2}$Rln$2$ between the two peaks.  Nearly the full Rln$2$ entropy is recovered for Na$_2$IrO$_3$ above the high temperature peak.  For materials in or close to the Kitaev QSL state a two-peak structure arises from the fractionalization of the spin into two kinds of Majorana Fermions \cite{Nasu2015, Yamaji2016}.  The entropy Rln$2$ is released in two equal parts leading to a plateau in $S_{mag}$ between the two peaks with a value ${1\over 2}$Rln$2$.  Thus our results provide the first thermodynamic evidence of proximity to the Kitaev QSL for Na$_2$IrO$_3$ materials.  Although results for Li$_2$IrO$_3$ are qualitatively similar, the quantitative agreement with theoretical predictions is not as strong.  Lastly, the low temperature $C$ for Li$_2$IrO$_3$ shows an unusual $T^2$ dependence suggesting $2$-dimensional magnetic excitations which could be confirmed in future inelastic scattering measurements on single crystals.   
         
\noindent
\emph{Acknowledgments.--} We thank the X-ray facility at IISER Mohali.  KM acknowledges UGC-CSIR India for a fellowship.  YS acknowledges DST, India for support through Ramanujan Grant \#SR/S2/RJN-76/2010 and through DST grant \#SB/S2/CMP-001/2013.

\end{document}